\documentclass{elsart}

\usepackage{amssymb}
\usepackage{graphicx}

\begin{document}

\begin{frontmatter}

\title{Degree correlation effect of bipartite network on personalized recommendation}

\author[1,2,3]{Jian-Guo Liu}, \ead{liujg004@ustc.edu.cn}
\author[1,2,3]{Tao Zhou}, 
\author[4]{Zhao-Guo Xuan}, 
\author[1]{Hong-An Che}, 
\author[1,2]{Bing-Hong Wang}, 
\author[1,2,3]{Yi-Cheng Zhang}

\address[1]{Research Center of Complex Systems Science,
University of Shanghai for Science and Technology, Shanghai 200093,
P R China}
\address[2]{Department of Modern Physics,
University of Science and Technology of China, Hefei 230026, P R
China}
\address[3]{Department of Physics, University of Fribourg, Fribourg
CH-1700, Switzerland}
\address[4]{Institute of Systems Engineering, Dalian
University of Technology, Dalian 116024, P R China}

\begin{abstract}
In this paper, by introducing a new user similarity index base on
the diffusion process, we propose a modified collaborative filtering
(MCF) algorithm, which has remarkably higher accuracy than the
standard collaborative filtering. In the proposed algorithm, the
degree correlation between users and objects is taken into account
and embedded into the similarity index by a tunable parameter. The
numerical simulation on a benchmark data set shows that the
algorithmic accuracy of the MCF, measured by the average ranking
score, is further improved by 18.19\% in the optimal case. In
addition, two significant criteria of algorithmic performance,
diversity and popularity, are also taken into account. Numerical
results show that the presented algorithm can provide more diverse
and less popular recommendatoins, for example, when the
recommendation list contains 10 objects, the diversity, measured by
the hamming distance, is improved by 21.90\%.
\end{abstract}


\end{frontmatter}

\section{Introduction}
With the expansion of Internet \cite{Broder2000} and widely
applications of {\it Web 2.0}, how to efficiently help people obtain
information that they truly need is a challenging task nowadays
\cite{Resnkck1997}. Recommender systems have become an effective
tool to address the information overload problem by predicting the
user's interests and habits based on their historical selections or
collections, which have been used to recommend books and CDs at
Amazon.com, movies at Netflix.com, and news at Versifi Technologies
(formerly AdaptiveInfo.com) \cite{Adomavicius2005}. Motivated by the
practical significance to the e-commerce and society, study of
recommender systems has caught increasing attentions and become an
essential issue in Internet applications such as e-commerce systems
and digital library systems \cite{Ecommerce2001}. A personalized
recommender system includes three parts: data collection, model
analysis and the recommendation algorithm, among which the algorithm
is the core part. Various kinds of algorithms have been proposed
thus far, including collaborative filtering approaches
\cite{Herlocker2004,Konstan1997,Liu2009,Liu2008b,LiuRR2009},
content-based analyses \cite{Balab97,Pazzani99}, hybrid algorithms
\cite{Pazzani1997,Basu1998,Good1999}, and so on. For a review of
current progress, see Refs. \cite{Adomavicius2005,Liu2009b} and the
references therein.

One of the most successful recommendation algorithms, called {\it
collaborative filtering} (CF), has been developed and extensively
investigated over the past decade
\cite{Herlocker2004,Konstan1997,Huang2004}. When predicting the
potential interests of a given user, CF algorithm firstly identifies
a set of similar users from the past records and then makes
predictions based on the weighted combination of those similar
users' opinions. Despite its wide applications, CF algorithm suffers
from several major limitations including system scalability and
accuracy \cite{Sarwar2000}. Therefore, the current CF algorithms
still require further improvements to make recommendations more
effective. Recently, some physical dynamics, including mass
diffusion \cite{Zhang2007b,Zhou2007} and heat conduction
\cite{Zhang2007a}, have found their applications in personalized
recommendations. Liu {\it et al.} \cite{Liu2009} introduced the mass
diffusion process to compute the user similarity of CF, and found
that the modified algorithm has remarkably higher accuracy than the
standard CF. In this method, all of the objects and users with far
different degrees have been treated equally, in other words, the
degree correlations between objects and users are neglected. For
example, suppose a user with small degree has collected a
small-degree object, the edge connecting them represents a very
special taste of the user, while the information contained in the
edges connecting an active user and a popular object is less
meaningful. Therefore, we argue that the user similarity index could
be improved by considering the degree correlation of the user-object
bipartite network. The numerical results show that the improved
index that depresses the influence of mainstream preferences can
provide more accurate and more diverse recommendations.

\section{Method}
Suppose there are $m$ objects and $n$ users in a recommender system.
Denote the object set as $O = \{o_1,o_2, \cdots, o_m\}$ and the user
set as $U$ = $\{u_1, u_2,$ $\cdots,$  $u_n\}$, a recommender system
can be fully described by an adjacent matrix $A=\{a_{ij}\}\in
R^{m,n}$, where $a_{ij}=1$ if $o_i$ is collected by $u_j$, and
$a_{ij}=0$ otherwise. In the standard CF, the user or object
similarities are calculated firstly, then the predictions are
computed accordingly. If $u_i$ has not yet collected $o_j$ (i.e.,
$a_{ji}=0$), the predicted score, $v_{ij}$, is given as
\begin{equation}\label{equation1}
v_{ij}=\frac{\sum_{l=1}^ns_{li}a_{jl}}{\sum_{l=1}^ns_{li}},
\end{equation}
where $s_{li}$ is the similarity between user $u_l$ and $u_i$. The
widest used similarity indices are the {\it Sorenson Index}
\cite{Soresen1948} and the {\it Salton Index} \cite{Salton1983},
however, they only rely on the users' degrees and the number of
common collected objects, without consideration of the influence of
degree correlation between users and objects. Inspired by the mass
diffusion process proposed by Zhou {\it et al.} \cite{Zhou2007}, Liu
{\it et al.} \cite{Liu2009} proposed a modified CF to improve the
algorithmic accuracy by using the mass diffusion process to compute
the user similarities, and they found that the diversity of
recommendations is also enhanced.
Although this algorithm has improved the standard CF, however, the
degree correlation between users and objects has not been
considered, thus every edge has the same contribution to the
diffusion process. If both of $u_i$ and $u_j$ have selected an
object $o_l$, they probably have similar tastes or interests.
Provided the degree of $o_l$ is very large (object $o_l$ is very
popular), this taste (the favor for $o_l$) is ordinary and it does
not mean $u_i$ and $u_j$ are very similar. Therefore, its
contribution to $s_{ij}$ should be weaken. On the other hand,
provided that a user $u_i$ with small degree has collected an
unpopular object $o_l$ (the degree of $o_l$ is very small), this
taste should be very special, the contribution of the edge
connecting $u_i$ and $o_l$ should be enlarged. It is not very
meaningful if a user with large degree has selected a popular
object, while if an unpopular object is selected by a small-degree
user, this edge would contain rich information on personalized
preference. Accordingly, the contribution of the edge connecting
$u_i$ and $o_l$ should be negatively correlated with $k(u_i)k(o_l)$.
We assume a certain amount of resource (e.g. recommendation power)
is associated with each user, and the weight $s_{ij}$ represents the
proportion of the resource $u_j$ would like to distribute to $u_i$.
Following a network-based resource-allocation process where each
user distributes his/her initial resource to all the objects he/she
has collected, and then each object sends back what it has received
to all the users who collected it, considering the correlation
between users and objects, the weight $s_{ij}$ (the fraction of
initial resource $u_j$ eventually gives to $u_i$) can be expressed
as
\begin{equation}\label{equation0}
s_{ij}=\frac{1}{k(u_j)}\sum^m_{l=1}\frac{a_{li}(k_{u_j}k_{o_l})^{\lambda}\cdot
a_{lj}(k_{u_i}k_{o_l})^{\lambda}}{k(o_l)},
\end{equation}
where $\lambda$ is a tunable parameter controlling the effect of
degree correlation. Based on the above definition, given a target
user $u_i$, the algorithm is given as following: (i) Calculating the
user similarity matrix $\{s_{ij}\}\in R^{n,n}$ based on the
diffusion process, as shown in Eq. (\ref{equation0}); (ii) For each
user $u_i$, according to Eq. (\ref{equation1}), calculating the
predicted scores for his/her uncollected objects; (iii) Sorting the
uncollected objects in descending order of the predicted scores, and
those objects in the top will be recommended.

\section{Algorithmic performance metrics}
To test a recommendation method on a dataset we randomly remove 10\%
of the links as the probe set and apply the algorithm to the
remainder (training set) to produce a recommendation list for each
user. We then employ three different metrics, one to measure
accuracy in recovery of deleted links and two to measure
recommendation popularity and diversity.

\subsection{Average ranking score} An accurate method will put preferable objects in higher places. The
average ranking score is adopted to measure the accuracy, which is
defined as follows. For an arbitrary user $u_i$, if the entry
$u_i$-$o_j$ is in the probe set (according to the training set,
$o_j$ is an uncollected object for $u_i$), we measure the position
of $o_j$ in the ordered list. For example, if there are $L_i=10$
uncollected objects for $u_i$, and $o_j$ is the 3rd from the top, we
say the position of $o_j$ is $3/10$, denoted by $r_{ij}=0.3$. Since
the probe entries are actually collected by users, a good algorithm
is expected to give high recommendations to them, leading to small
$r_{ij}$. Therefore, the mean value of the position, $\langle
r\rangle$, averaged over all the entries in the probe, can be used
to evaluate the algorithmic accuracy: the smaller the average
ranking score, the higher the algorithmic accuracy, and vice verse.

\begin{figure}[ht]
\center\scalebox{0.35}[0.35]{\includegraphics{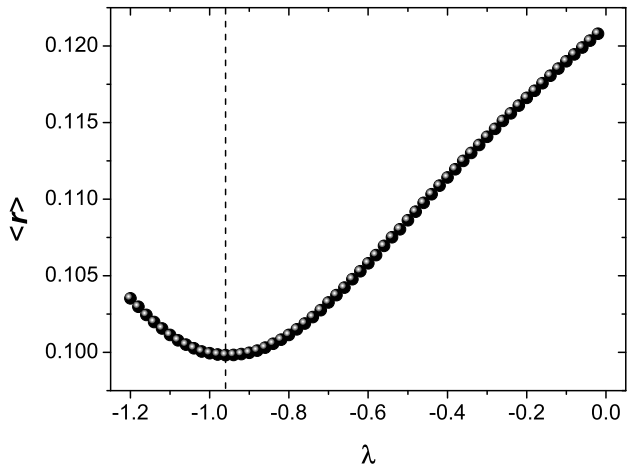}}
\caption{The average ranking score $\langle r\rangle$ vs. $\lambda$
for the algorithm. The optimal $\lambda_{\rm opt}$, corresponding to
the minimal $\langle r\rangle= 0.0998$, is $\lambda_{\rm
opt}=-0.96$. When $\lambda=0$, the algorithm degenerates to the
accuracy of the CF based on the diffusion process. All the data
points are averaged over ten independent runs with different
data-set divisions.}\label{Fig1}
\end{figure}

\subsection{Popularity and diversity}
Besides accuracy, the average degree of all recommended objects,
$\langle k\rangle$, and the mean value of Hamming distance, $S$, are
taken into account to measure the algorithmic popularity and
diversity \cite{Zhou2007b}. The smaller average degree,
corresponding to the less popular objects, are preferred since those
lower-degree objects are hard to be found by users themselves. For
example, suppose there are 10 perfect movies not yet known for user
$u_i$, 7 of which are widely popular, while the other three fit a
certain specific taste of $u_i$. An algorithm recommending the 7
popular movies is very nice for $u_i$, but he may feel even better
about a recommendation list containing those two unpopular movies.
In addition, the personalized recommendation algorithm should
present different recommendations to different users according to
their tastes and habits. The diversity can be quantified by the
average Hamming distance, $S=\langle H_{ij}\rangle$, where
$H_{ij}=1-Q_{ij}/L$, $L$ is the length of recommendation list and
$Q_{ij}$ is the overlapped number of objects in $u_i$ and $u_j$'s
recommendation lists. The largest $S=1$ indicates the
recommendations to all of the users are totally different, in other
words, the system has highest diversity. While the smallest $S=0$
means that the recommendations for different users are exactly the
same.

\begin{figure}[ht]
\center\scalebox{0.35}[0.35]{\includegraphics{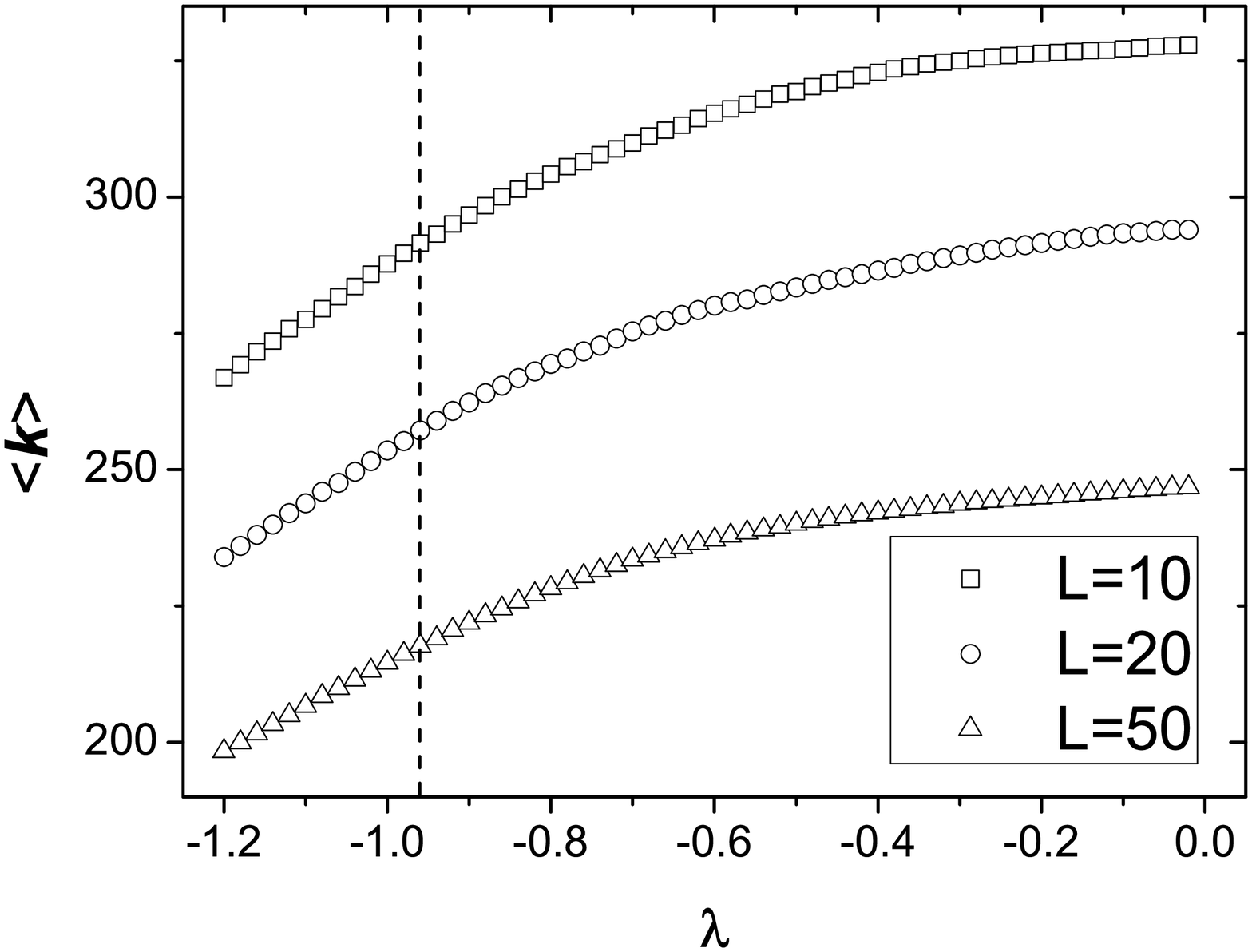}}
\caption{Average degrees of recommended objects, $\langle k\rangle$
vs. $\lambda$. Squares, circles and triangles represent lengths
$L=10, 20$ and $50$, respectively. All the data points are averaged
over ten independent runs with different data-set
divisions.}\label{Fig2}
\end{figure}

\begin{figure}[ht]
\center\scalebox{0.35}[0.35]{\includegraphics{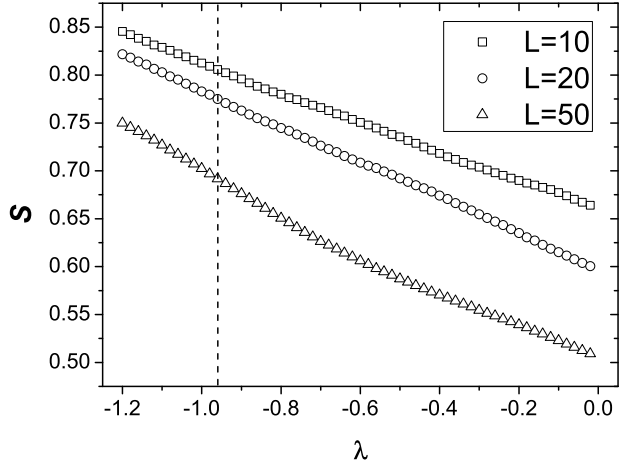}}
\caption{The diversity $S$ vs. $\lambda$. Squares, circles and
triangles represent the lengths $L=10, 20$ and $50$, respectively.
All the data points are averaged over ten independent runs with
different data-set divisions.}\label{Fig3}
\end{figure}


\section{Numerical results}
A benchmark dataset, namely MovieLens, is used to test the above
algorithm, which consists of 1682 movies (objects) and 943 users.
The users vote movies by discrete ratings from one to five. We
applied a coarse-graining method: A movie is set to be collected by
a user only if the given rating is larger than 2. The original data
contains $10^5$ ratings, 85.25\% of which are larger than 2, that
is, the user-object (user-movie) bipartite network after the coarse
gaining contains 85250 edges.

Figure \ref{Fig1} reports the algorithmic accuracy as a function of
$\lambda$. The curve has a clear minimum around $\lambda=-0.96$,
which strongly support the above discussion that to depress the
influence of the users or objects with large degrees could enhance
the accuracy. Compared with the routine case ($\lambda=0$), the
average ranking score can be reduced by 18.19\% at the optimal case,
which indeed a great improvement. Figure \ref{Fig2} reports the
average degree of all recommended movies as a function of $\lambda$.
When $\lambda<0$ the average object degree is positively correlated
with $\lambda$, thus to depress the influence of edges connecting
active users and popular objects gives more opportunity to the
unpopular objects, which is consistent with our expectation. Figure
\ref{Fig3} exhibits a negative correlation between $S$ and
$\lambda$, indicating that to depress the influence of the edges
connecting active users and popular objects makes the
recommendations more personalized. When $L=10$, the diversity $S$ is
increased from 0.661 (corresponding to the case when $\lambda=0$) to
0.806 (corresponding to the optimal case $\lambda=-0.96$), improved
by 21.90\%.

\section{Conclusions and discussions}
In this paper, a modified collaborative filtering algorithm is
presented to improve the algorithmic accuracy by depressing the
influence of edges connecting active users and popular objects. The
algorithmic accuracy, measured by the average ranking score, can be
improved by 18.19\%. Beside accuracy, two significant criteria of
algorithmic performance, popularity and diversity, are also taken
into account. A good recommendation algorithm should not only has
higher accuracy, but also help the users uncovering the hidden
information, corresponding to those objects with low degrees.
Therefore, the average object degree is a meaningful measurment for
a recommendation algorithm. In addition, a personalized recommender
system should provide each user personalized recommendations
according to his/her interests and habits, therefore, the diversity
of recommendations plays a crucial role to quantify the
personalization. The numerical results show that the presented
algorithm outperforms the standard CF in all three criteria,
accuracy, popularity and diversity.


Since the power computation takes much more time than
multiplication, this algorithm would take longer time to get the
user similarities. Throughout the numerical simulation results, we
could find that the optimal $\lambda_{\rm opt}$ is close to -1. When
$\lambda=-1$, the corresponding $\langle r\rangle=0.0995$, which is
also improved 18.07\%, and the average object degree and diversity
are getting even better, where the diversity $S$ has been improved
by 23.40\%. Therefore, in real application, the parameter could be
set as -1, which ensures that the algorithmic complexity is as same
as a parameter free CF.

How to automatically find out relevant information for diverse users
is a long-standing challenge in the modern information science, the
presented algorithm also could be used to find the relevant
reviewers for the scientific papers or funding applications
\cite{Liu1,Liu2}, and the link prediction in social and biological
networks \cite{Zhou2009}. We believe the current work can enlighten
readers in this promising direction.

We acknowledge {\it GroupLens} Research Group for providing us the
data set. This work is partially supported by SBF (Switzerland)
project C05.0148 (Physics of Risk), the National Natural Science
Foundation of China (Grant Nos. 10635040 and 60744003), the Swiss
National Science Foundation (project 205120-113842), the Specialized
Research Fund for the Doctoral Program of Higher Education of
China.(Grant No. 20060358065), and by the Research Fund of the
Education Department of Liaoning of China (20060140).

\end{document}